\documentclass[aps,prd,twocolumn,superscriptaddress,amsfont,graphicx,nofootinbib,preprintnumbers]{revtex4}%
\usepackage{color,graphicx,epsfig}
\usepackage{ifpdf}
\usepackage{amsmath}
\usepackage{bm}
\usepackage{color}
\usepackage[english]{babel}
\usepackage{graphicx}%
\usepackage{amsfonts}%
\usepackage{amssymb}
\usepackage{braket}
\usepackage{hyperref}

\definecolor{nicered}{rgb}{0.7,0.1,0.1}
\definecolor{nicegreen}{rgb}{0.1,0.5,0.1}
\hypersetup{colorlinks,citecolor= nicegreen,linkcolor= nicered}

\newcommand{\slashed}{\slash \hspace{-0.19cm}}

\newcommand{\beq}{\begin{equation}}
\newcommand{\eeq}{\end{equation}}
\newcommand{\bea}{\begin{eqnarray}}
\newcommand{\eea}{\end{eqnarray}}

\begin{document}

\def\LjubljanaFMF{Faculty of Mathematics and Physics, University of Ljubljana,
 Jadranska 19, 1000 Ljubljana, Slovenia }
\def\LjubljanaIJS{Jo\v zef Stefan Institute, Jamova 39, 1000 Ljubljana, Slovenia}
\def\CERN{CERN, Theory Division, CH-1211 Geneva 23, Switzerland}
\def\MIT{Center for Theoretical Physics, Massachusetts Institute of Technology, Cambridge, Massachusetts 02139, USA}
\def\Cincy{Department of Physics, University of Cincinnati, Cincinnati, Ohio 45221,USA} 

\title{
Lepton flavor universality violation without new sources of quark flavor violation
}

\author{Jernej F.\ Kamenik} 
\email[Electronic address:]{jernej.kamenik@cern.ch} 
\affiliation{\LjubljanaIJS}
\affiliation{\LjubljanaFMF}

\author{Yotam Soreq} 
\email[Electronic address:]{soreqy@mit.edu} 
\affiliation{\MIT}

\author{Jure Zupan} 
\email[Electronic address:]{zupanje@ucmail.uc.edu} 
\affiliation{\CERN}
\affiliation{\Cincy}

\preprint{CERN-TH-2017-089, MIT-CTP/4904}

\date{\today}
\begin{abstract}
We show that new physics models without new flavor violating interactions can explain the recent anomalies in the $b\to s\ell^+\ell^-$ transitions. The $b\to s\ell^+\ell^-$ arises from a $Z'$ penguin which automatically predicts the $V-A$ structure for the quark currents in the effective operators. This framework can either be realized  in a renormalizable $U(1)'$ setup or be due to new strongly interacting dynamics. The dimuon resonance searches at the LHC are becoming sensitive to this scenario since the $Z'$ is relatively light, and could well be discovered in future searches by ATLAS and CMS. 
\end{abstract}

\maketitle

{\bf Introduction.}
Lepton flavor universality~(LFU) of electroweak interactions is one of the key predictions of the standard model~(SM). The electric charge is copied from one generation of fermions to the other, so that the photon couples with the same strength to the electron as it does to the muon and the tau lepton. Similarly, the $Z$  boson couples in the same way to all three generations of leptons, a fact that has been tested at the permille level for on-shell $Z$ couplings at LEP~\cite{Olive:2016xmw,ALEPH:2005ab}. Any deviation from LFU either in on-shell processes or from off-shell exchanges would be a clear indication of new physics~(NP) (LFU violations from differing charged lepton masses are usually negligibly small, but will be kept in our discussion when needed).

In the past several years a number of measurements of the $b\to s\ell^+\ell^-$ transitions \cite{Aaij:2014pli,Aaij:2014ora,Aaij:2015esa,Aaij:2015oid,Wehle:2016yoi,Abdesselam:2016llu,ATLAS-CONF-2017-023,CMS-PAS-BPH-15-008,Bifani-talk} have been showing a pattern of deviations from the SM predictions \cite{Altmannshofer:2014rta,Descotes-Genon:2015uva,Hurth:2016fbr,Altmannshofer:2017fio,Ghosh:2014awa} (for most recent global fits see \cite{Geng:2017svp,Ciuchini:2017mik,Hiller:2017bzc,DAmico:2017mtc,Altmannshofer:2017yso,Capdevila:2017bsm}).  While none of the deviations by themselves is yet statistically significant, and some of them require precise control of hadronic uncertainties, it is quite striking that the deviations appear also in such clean observables as the ratios that probe LFU.
Currently there is a 2.6$\sigma$ discrepancy with the SM in $R_K=\big(d\Gamma(B\to K\mu^+\mu^-)/dq^2\big)/\big(d\Gamma(B\to K e^+e^-)/dq^2\big)$~\cite{Aaij:2014ora},
\beq
R_{K,[1,6]{\rm GeV^2}}=0.745\pm0.090,
\eeq
and a $2.2-2.5\sigma$ discrepancy in the related mode with the vector meson, $R_{K^*}=\big(d\Gamma(B\to K^*\mu^+\mu^-)/dq^2\big)/\big(d\Gamma(B\to K^* e^+e^-)/dq^2\big)$~\cite{Bifani-talk},
\beq
\begin{split}
R_{K^*,[0.045,1.1]{\rm GeV^2}}&=0.66^{+0.11}_{-0.07} \, , 
\\
R_{K^*,[1.1,6]{\rm GeV^2}}&=0.69^{+0.12}_{-0.08} \, .
\end{split}
\eeq
If confirmed, these would constitute a discovery of NP. 

The NP models that have been put forward to explain the $b\to s \ell^+\ell^-$ anomalies fall into two categories. Most of the analyses so far have focused on the case where the $b\to s \ell^+\ell^-$ transition receives a contribution from a tree level exchange of a new heavy vector boson, $Z'$, with flavor violating couplings to $b$ and $s$ quarks, as well as couplings to either electrons \cite{Carmona:2015ena} or muons~\cite{Descotes-Genon:2013wba,Buras:2013qja,Gauld:2013qja,Buras:2013dea,Altmannshofer:2014cfa,Celis:2015ara,Falkowski:2015zwa,Crivellin:2015lwa,Crivellin:2015mga,Boucenna:2016qad,Crivellin:2016ejn,Megias:2016bde,Megias:2017ove,GarciaGarcia:2016nvr}  (in the case of Ref.~\cite{Belanger:2015nma}  the latter is generated at loop level),  or through tree level exchange of leptoquarks \cite{Hiller:2014yaa,Gripaios:2014tna,Becirevic:2015asa,Varzielas:2015iva,Fajfer:2015ycq,Becirevic:2016yqi}. The other set of models generates the $b\to s \ell^+\ell^-$ through box loop diagrams with new heavy fields \cite{Gripaios:2015gra,Arnan:2016cpy}. 
Both of these sets of solutions require flavor changing couplings beyond those present in the SM. One thus needs to make sure that the generated flavor changing transitions are consistent with other precision flavor observables such as $B_s-\bar B_s$, $D-\bar D$ mixing, etc.

\begin{figure}[!t]
\centering
\includegraphics[width=0.48\textwidth]{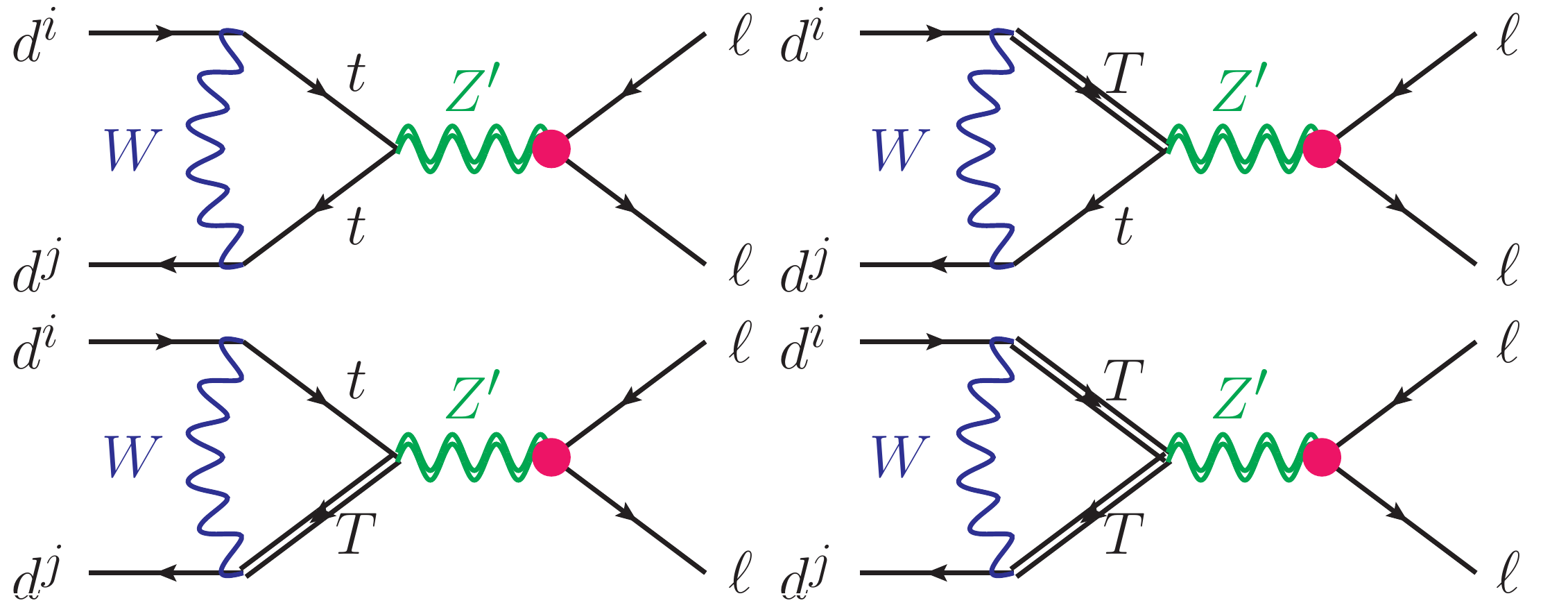}
\caption{The NP contributions to the $d^i\to d^j \ell\ell$ processes from the exchange of a $Z'$ that couples to the top quark and a heavy top partner $T$.}
\label{fig:Feynman:flavor}
\end{figure}

In this paper we show that there is a third class of models where all the NP couplings are flavor diagonal -- but not flavor universal. The simplest realization is in terms of a $Z'$ whose dominant couplings in the SM sector are to the right-handed top quarks and to the muons, see Fig.~\ref{fig:Feynman:flavor}. Other realizations are possible, for example in strongly coupled scenarios as we briefly discuss below.  

 The NP models that we are proposing as possible explanations of $b\to s \ell^+\ell^-$ anomalies have several salient features. They are examples of NP with (general) minimal flavor violation (MFV) \cite{Chivukula:1987py,Hall:1990ac,Buras:2000dm,DAmbrosio:2002vsn,Kagan:2009bn} and thus easily satisfy the present experimental bounds from other flavor changing neutral current transitions, beside $b\to s \ell^+\ell^-$. The $b\to s \ell^+\ell^-$ transition is generated via the exchange of the SM $W$ gauge boson in the loop. This class of models thus leads automatically to the $V-A$ structure of the quark current in the NP operators, as preferred by the global fits to the data~\cite{Altmannshofer:2014rta,Descotes-Genon:2015uva,Hurth:2016fbr,Altmannshofer:2017fio}. There is more freedom in the structure of couplings to muons, where both $V-A$ and $V+A$ currents are possible. Finally, since in this class of models the $b\to s \ell^+\ell^-$ transition is generated at the one-loop level, the $Z'$ is quite light, with a mass of a few hundred GeV and can be searched for at the LHC in high $p_T$ processes. 

{\bf General discussion.}
The effective weak Hamiltonian that describes the $b\to s\ell^+\ell^-$ transitions is given by 
\beq
{\cal H}_{\rm eff}=-\frac{4 G_F}{\sqrt 2} V_{tb} V_{ts}^* \frac{e^2}{16\pi^2}\sum_i\big( C_i^\ell O_i^\ell +C_i'{}^\ell O_i'^\ell\big)+{\rm H.c.},
\eeq
where $e$ is the EM gauge coupling and the sum runs over the dimension-five and dimension six-operators. 
Denoting SM and NP contributions to the Wilson coefficients as $C_i^\ell = C_i^{\ell,\rm SM} + C_i^{\ell,\rm NP}$,  global analyses of all $b\to s\ell^+\ell^-$ indicate a nonvanishing $C_9^{\mu,\rm NP}$, with some preference for a NP solution with  $C_9^{\mu,\rm NP}=-C_{10}^{\mu,\rm NP} \simeq 0.60(15)$; see, e.g., \cite{Altmannshofer:2017fio}. Here  the relevant four-fermion operators are $O_9^\ell=\big(\bar s\gamma_\mu P_Lb\big) \big(\bar \ell \gamma^\mu \ell\big)$,  and $O_{10}^\ell=\big(\bar s\gamma_\mu P_L b\big) \big(\bar \ell \gamma^\mu \gamma_5 \ell\big)$. The data thus imply the presence of NP contributions with a $V-A$ structure in the quark sector. However, additional contributions of comparable magnitude but with a $V+A$ structure from the NP operators $O_9'^\ell=\big(\bar s\gamma_\mu P_R b\big) \big(\bar \ell \gamma^\mu \ell\big)$,   $O_{10}'^\ell=\big(\bar s\gamma_\mu P_R b\big) \big(\bar \ell \gamma^\mu \gamma_5 \ell\big)$ are still allowed by the current data. 

In the class of models we are considering only the 
$O_9^\ell$ and $O_{10}^\ell$ are generated at one loop, see Fig. \ref{fig:Feynman:flavor}. The $V-A$ current in the quark sector is a clear prediction of the models, while the structure of the couplings to leptons depends on the details of the model. For simplicity we assume that NP  predominantly affects the $b\to s\mu^+\mu^-$ transition and not the $b\to s e^+e^-$. This leads to LFU violation when comparing $b\to s \mu^+\mu^-$ with $b\to s e^+e^-$. It also modifies the total rates in various $b\to s \mu^+\mu^-$ decays, in accordance with indications of global fits \cite{Altmannshofer:2014rta,Descotes-Genon:2015uva,Hurth:2016fbr,Altmannshofer:2017fio}. On the other hand $B_s$, $B_d$ and $K^0$ mixing via $Z'$ exchange arises only at the two-loop level and is well within present experimental and theoretical precision.

Since the NP sector does not contain new sources of flavor violation, this class of models respects the MFV ansatz. 
In MFV, a shift to $C^\ell_{9,10}$ can be correlated with the analogue contributions to rare kaon decays. For instance, the $K^+ \to \pi^+ \nu\bar \nu (\gamma)$ decay branching ratio is modified to~\cite{Buras:2015qea}\footnote{This is for leptons in an isospin singlet state, while for an isospin triplet combination, the NP contribution flips its sign.} 
\begin{align}
\mathcal B(K^+ \to \pi^+ \nu\bar \nu (\gamma)) = (8.4\pm 1.0)\times 10^{-11}   \nonumber \\
\times\frac{1}{3} \sum_\ell \left| 1+ \frac{s_W^2  (C^{\ell,\rm NP}_9 - C^{\ell,\rm NP}_{10})}{X_{\rm SM}}  \right|^2 \,,
\end{align}
where $X_{\rm SM} = X_t + (X_c + \delta X_{c,u}) V_{us}^4 V_{cs} V_{cd}^* / V_{ts} V_{td}^* \simeq 2.10 + 0.24 i$ with $X_i$ defined, e.g., in \cite{Brod:2010hi}, and have written for  the weak mixing angle $s_W \equiv \sin \theta_W \simeq 0.48$, $c_W \equiv \cos \theta_W$. For values of $C_{9,10}^{\mu,\rm NP}$ that are preferred by current $b \to s \ell\ell$ data, the resulting effect in $K\to \pi \nu\bar\nu$ is small compared to current experimental uncertainties, but could be within reach of the ongoing NA62 experiment~\cite{Anelli:2005ju}. Similar comments apply to the theoretically very clean $K_L\to \pi^0 \nu\bar \nu$ decay. The decay $K_L\to \pi^0 \mu^+\mu^-$ is modified at the level of ${\mathcal O}(5\%)$ by such NP models. To observe these effects the experimental sensitivity \cite{AlaviHarati:2000hs} would need to be improved by two orders of magnitude in conjunction with some improvements in theoretical precision \cite{Mescia:2006jd}. The decay modes $K^+\to \pi^+ e^+ e^-$ and $K^+\to \pi^+ \mu^+ \mu^-$ are dominated by long distance contributions, while the NP contributions are expected to only give effects below the permille level and thus be unobservable. The same is true for the $K_L \to \mu^+\mu^-$ transition, where again the NP contribution is drowned by the SM long distance effects.

{\bf The minimal aligned $U(1)'$ model.}
We discuss next the simplest realization of the above framework. We restrict ourselves to the case where on the leptonic side only the muons are affected by NP. The minimal model has a new 
$U(1)'$ gauge symmetry that is spontaneously broken through the vacuum expectation value~(VEV) of a scalar field, $\Phi$, transforming as  $\Phi \sim (1,1,0,q')$ under $SU(3)_C \times SU(2)_L \times U(1)_Y \times U(1)'$. The model contains, in addition, a colored Dirac fermion $T' \sim (3,1,2/3,q')$. The SM is thus supplemented by the Lagrangian 
\beq
\begin{split}
\mathcal L_{U(1)'} =& |(D_\mu \Phi)|^2 - \frac{m^2_\phi}{2 \tilde v^2} \Big(\Phi^2 - \frac{\tilde v^2}{2}\Big)^2 \\
&+ \bar T' (i \slashed D - M_{T}) T' - \frac{1}{4} F^{\prime 2}_{\mu\nu} \,,
\end{split}
\eeq
where $D_\mu \supset i \tilde g q'  Z'_\mu$, the $U(1)'$ part of the covariant derivative, 
$F'_{\mu\nu} = \partial_\mu Z'_\nu-  \partial_\nu Z'_\mu$ the field strength for the gauge boson $Z'$, and $\Phi = (\phi + \tilde v)/\sqrt 2$.  Here $\tilde g$ is the $U(1)'$ gauge coupling, $\tilde v$ is the VEV that breaks the $U(1)'$, while $\phi$ is the physical scalar boson that obtains mass $m_\phi$ after spontaneous breaking of $U(1)'$.

\begin{figure}
\centering
\includegraphics[width=0.38\textwidth]{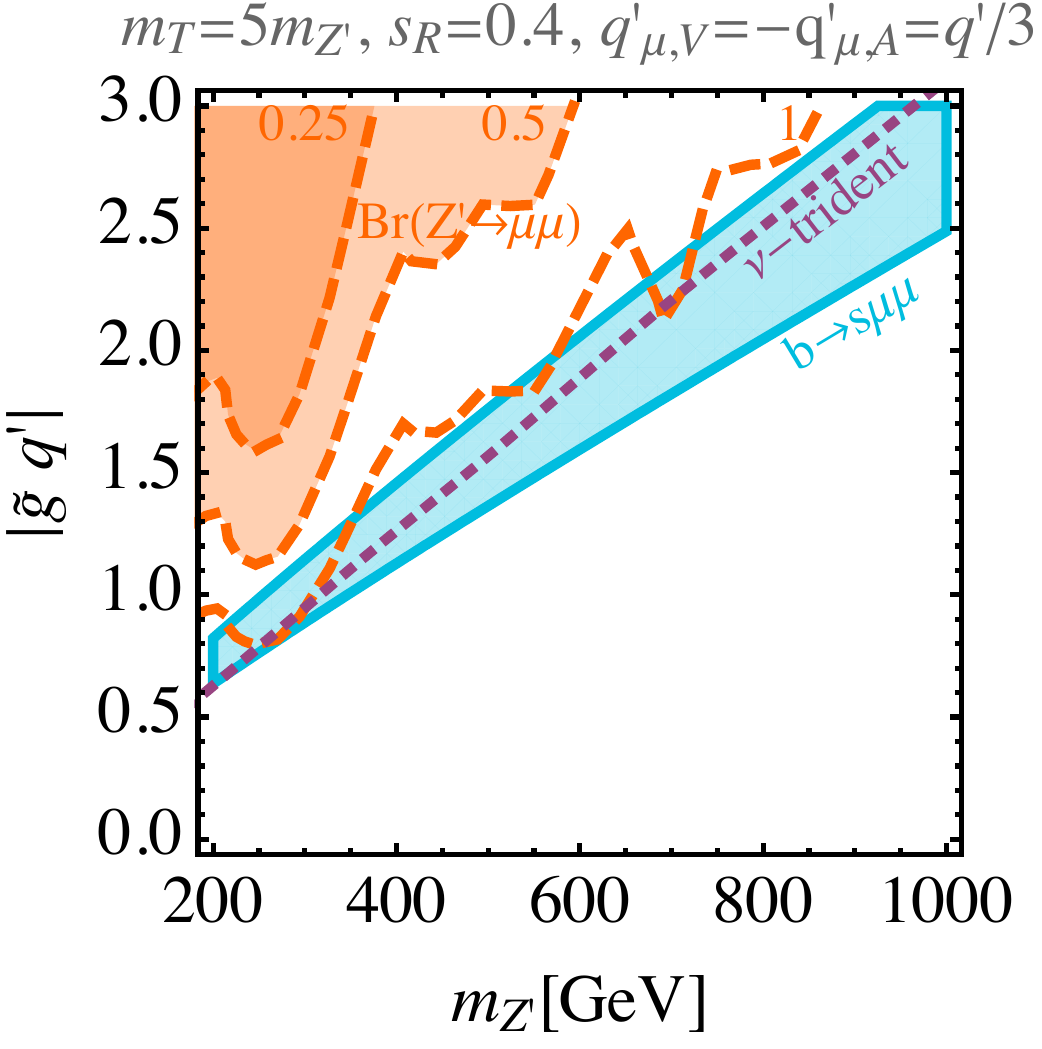}
\caption{The constraints in the $\tilde g$, $m_{Z'}$ plane coming from dimuon searches at the LHC for ${\rm Br}(Z'\to \mu^+\mu^-)=0.25, 0.50, 1$ (from darker to lighter orange). 
The area above the dashed purple line is $\nu$-trident production. 
The blue region shows the parameter space preferred by the $b\to s \ell^+\ell^-$ anomalies. }
\label{fig:param_space}
\end{figure}

All the SM fields are singlets under $U(1)'$. There are only three renormalizable interactions between the SM and the $U(1)'$ sector: the Higgs portal coupling $\Phi$ to the SM Higgs, $H$; the $U(1)'$ kinetic mixing with the SM hypercharge, $B_{\mu\nu}$; and a Yukawa-type coupling of $T$ and $\Phi$ to the SM right-handed up-quarks $u_R^i$,
\beq
\mathcal L_{\rm mix} = - \lambda' |\Phi|^2 |H|^2  - \epsilon B^{\mu\nu} F'_{\mu\nu} - (y^i_T \bar T' \Phi u^i_R + \rm H.c. )\,.
\eeq
The summation over generation index $i=1,2,3,$ is implied. While $y_T^i$ can in general take any values, we assume it is aligned with the right-handed up-quark  Yukawa coupling,  i.e., that the two satisfy the basis independent condition $\big[ y_T^\dagger y_T  \, , \ y_u^\dagger y_u \big] =0$.
In the up-quark mass basis thus $y^{ij}_u\sim {\rm diag}(0,0,y_t)$, and  $y_T^i\sim (0,0,y_T^t)$ so that at leading order $Z$($Z'$)-couplings to light quarks remain exactly SM-like (vanish); see Refs.~\cite{Aguilar-Saavedra:2013qpa, Fajfer:2013wca} for more detailed discussion. Such a structure is natural in flavor models of quark masses  where the commutator above does not vanish exactly, but it is still sufficiently small to avoid dangerous $Z-$ and $Z'-$mediated flavor changing neutral currents. 
For example, in Froggatt-Nielsen type models with horizontal $U(1)$  symmetry~\cite{Froggatt:1978nt} one has $y_T\sim y^y_T(c_t \lambda_C^3, c_c \lambda_C,1)$, with $\lambda_C\sim0.2$ and $c_{u,c}\sim{\mathcal O}(1)$. If $U(1)$ is gauged, the charm mixing~\cite{Bazavov:2017weg} bounds the corresponding $Z'$ to $m_{Z'}\gtrsim  |\text{Re}(c_u c_c^*)| \times 250\text{~GeV},$ for ${\mathcal O}(1)$ gauge couplings and large mixing between $t$ and $T$, as in Fig.~\ref{fig:param_space}. While these start to probe interesting parameter space they do not yet exclude the above explanation of the $b\to s \mu^+\mu^-$ anomaly.

In the rest of the paper we ignore the mixing of $T$ with the first two generations of quarks. For simplicity we also assume that $|y_T^t| \gg \lambda', \epsilon$, and neglect the Higgs portal and the kinetic mixing couplings.
After electroweak symmetry breaking the $t-T'$ part of the mass matrix, $\mathcal M_u$, for up-type quarks and $T'$ 
is given by
\beq
\mathcal M^{t-T'}_u = \left( \begin{array}{cc}  
y_t v/\sqrt{2} & 0 \\
y^t_T \tilde v / \sqrt{2} & M_T
\end{array} \right)\,,
\eeq
where $v \simeq 246$\,GeV is the SM electroweak~(EW) VEV. The mass eigenstates, $t, T$, with masses $m_t\simeq173$ GeV and $m_T$, are an admixture of the interaction states with the mixing angles for two chiralities, $\theta_{L,R}$, given by 
\begin{align}
 \tan (2\theta_{L} ) &= \frac{ y_t y_T^t  v \tilde v  }{ M_T^2 - (y_t v)^2/2 + (y^t_T \tilde v)^2/2}, \\
 \tan(2 \theta_R) &=  \frac{\sqrt{2}  y_T^t M_T  \tilde v  }{M_T^2 - (y_t v)^2/2 - (y^t_T \tilde v)^2/2}\,.
 \end{align}
 In the phenomenological analysis we will take $y_t v, y_T^t\tilde v\ll M_T$, in which case $\theta_R\sim y_T^t \tilde v/M_T$ and $\theta_L\sim \theta_R v/M_T$.
 The two mass eigenstates, $t, T$, have masses $m_{t}\simeq y_t v/\sqrt{2}$, $m_T\simeq M_T$, or more precisely,
 \begin{align}
 m_t m_T & =  M_T  \frac{y_t v}{\sqrt 2}  \,, \\
 2(m_t^2 + m_T^2)  &= 2 M_T^2 + (y_t v)^2 + (y^t_T \tilde v)^2.
 \end{align}
The couplings to the massive gauge bosons are thus given by
\begin{align}
	\mathcal L_{\rm int} 
&	= - \frac{g}{\sqrt 2} V_{ti} \left[ (c_{L} \bar t + s_{L} \bar T ) \slashed W^+   P_L d_i\right] + {\rm H.c.} \nonumber \\
&	 - \frac{g}{2 c_W} \left[ (c_{L} \bar t + s_{L} \bar T)  \slashed Z P_L  (c_{L}  t + s_{L}  T)   - 2 s_W^2 J^\mu_{\rm EM} Z_{\mu} \right] \nonumber \\
& 	- {\tilde g q'}{} \left[  (s_{L} \bar t - c_{L} \bar T)  \slashed Z' P_L  (s_{L}  t - c_{L}  T) \right. \nonumber \\
&	\qquad\qquad \left. + (s_{R} \bar t - c_{R} \bar T) \slashed Z' P_R (s_{R}  t - c_{R}  T)  \right]  \,,
\end{align}
where we used for shortness $s_{L,R}(c_{L,R})\equiv\sin\theta_{L,R}(\cos\theta_{L,R})$. The SM weak coupling constant is $g \equiv 2 m_W \sqrt{G_F \sqrt 2} \simeq 0.65$, $V_{ij}$ are the elements of the unitary $3\times 3$ CKM matrix, and $J_{\rm EM}^\mu \equiv 2 (\bar t \gamma^\mu t + \bar T \gamma^\mu T)/3$ is the relevant EM current.

In the limit $M_T \gg v,\tilde v$ the dominant effect is in the new $\bar t \slashed Z'  P_R t$ interaction which is suppressed by $1/M_T^2$, while modifications of $W$ and $Z$ couplings appear at ${\mathcal O}(1/M_T^4)$. The mixing angle $\theta_{L}$ is constrained by electroweak precision tests. The modification of the $\rho$ parameter is given by~\cite{Fajfer:2013wca}
\begin{align}
	\Delta \rho 
=&	 \frac{\alpha N_C}{16\pi s_W^2} \frac{m_t^2}{m_W^2} s_{L}^2 \bigg[  -(1+c_{L}^2) + s_{L}^2 r  \nonumber \\ 
&	\qquad\qquad+ 2 c_{L}^2 \frac{r}{r-1} \log r \bigg] + \mathcal O\left( \frac{m_Z^2}{m_T^2} \right)\,,
\end{align}
where $r\equiv m_T^2/m_t^2$. A comparison with the experimental value $\Delta \rho_{\rm exp} = (4 {}^{+3}_{-4}) \cdot 10^{-4}$~\cite{Olive:2016xmw} yields $s_{L} \lesssim 0.2 $ for $m_T \gtrsim 1$ TeV\,.

The renormalizable vector and axial muonic current couplings to $Z'$ are in general given by 
\beq
\mathcal L_{\rm eff}^{(\mu)}= -\tilde g \bar\mu \slashed Z' (q'_{\ell,V} + q'_{\ell,A} \gamma_5) \mu\,.
\eeq
We assume that the $Z'$ couplings to charged leptons are flavor diagonal and focus on couplings to muons. The effective couplings of $Z'$ to the muon, $q'_{\ell,V}, q'_{\ell,A}$, depend on the embedding of $U(1)'$ in the UV theory. 
For instance, if only $\mu_L$ couples to $Z'$, then $q_{\ell,V}'=-q_{\ell,A}'$, giving $C_9^{\mu,{\rm NP}}=-C_{10}^{\mu,{\rm NP}}$. This possibility is somewhat preferred by present $b\to s\ell\ell$ global fits. 
Such structure arises, if the SM muon EW doublet, $L=(\mu_L,\nu_\mu)$, mixes with a heavy Dirac fermion lepton, $L_T$, through a Yukawa interaction $y_\mu \bar L \Phi L_T$ (a possibility of this type was first discussed in \cite{Sierra:2015fma}). The $L_T$ has the same electroweak charges as $L$, but is in addition charged under the $U(1)'$ with the opposite charge to $\Phi$. The $L_T$ decays predominantly through $L_T\to \mu Z, \nu W \to \mu \nu \bar \nu$. Chargino searches at the LHC in the dilepton+MET channel bound  $M_{L_T}\gtrsim 600$ GeV from $L_T$ pair production \cite{ATLAS-CONF-2016-096,Khachatryan:2014qwa}.  If there is in addition a heavy $U(1)'$ lepton with electroweak charges of the right-handed muon then there is no fixed relation between $C_9^{\mu,{\rm NP}}$ and $C_{10}^{\mu,{\rm NP}}$.  
Furthermore, $Z'$ can also couple to electrons and taus, a possibility we do not pursue in detail, but may be important for LHC searches and their relation to LFU violating observables in $B$ decays, as well as to $K\to \pi \nu\bar\nu$ decays. Depending on the details of how the leptonic sector is extended one may also potentially explain the $g-2$ anomaly.

The leading $Z'$ effects in rare semileptonic $B$ meson decays are captured by the shifts to the Wilson coefficients (see also \cite{Aebischer:2015fzz})
\beq
C^{\mu,\rm NP}_{9,10} = \frac{1}{2}q' q^{\prime }_{\mu,V,A} \frac{m_t^2}{m_{Z'}^2} \frac{\tilde g^2}{e^2} s_R^2 \log \left( \frac{m_T^2}{m_W^2} \right) + \ldots\,,
\eeq
where we have kept only the dominant, logarithmically enhanced term. We observe that sufficiently large $C_{9,10}^{ \mu,\rm NP}$ as preferred by current data can be generated for $\mathcal O(1)$ values of the $U(1)'$ charges and gauge coupling, provided $m_{Z'}$ lies below $\mathcal O (\rm TeV)$; see Fig.~\ref{fig:param_space}\,.

\begin{figure}[!t]
\includegraphics[width=0.48\textwidth]{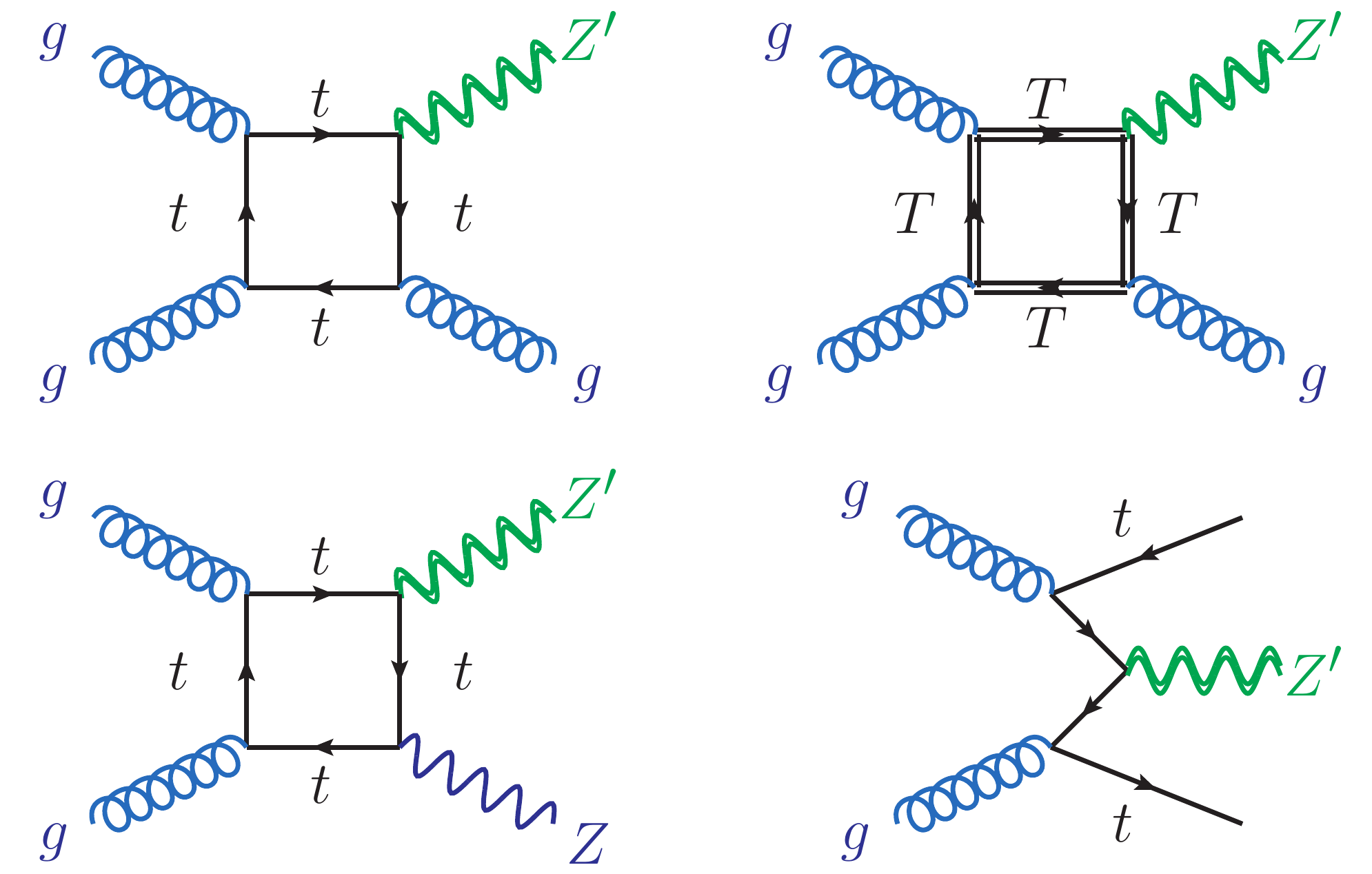}
\caption{Representative Feynman diagrams for $p p\to j Z'$ production at the LHC (first row, in addition to the box diagrams there are also triangle diagrams), and for $pp\to Z Z'$ and $pp\to t\bar t Z'$ production. }
\label{fig:Feynman:LHC}
\end{figure}

The searches at the LHC for dimuon resonances could put important bounds on the $Z'$ couplings and its mass or lead to its discovery~\cite{ATLAS-CONF-2017-027,Khachatryan:2016zqb}. 
The most important production channels are the tree level $p p \to t \bar t Z'$, as well as $p p \to Z Z' $ and $p p \to j Z' $ production through top and $T$ loops. The representative diagrams for these are shown in Fig.~\ref{fig:Feynman:LHC} (see also~\cite{Greiner:2014qna}). For the calculation we use {\tt MadGraph5\_aMC@NLO}~\cite{Alwall:2014hca} with a modified model file~\cite{Fuks:2017vtl} for the model of Ref.~\cite{Altarelli:1989ff}.

The $Z'$ boson decays to pairs of muons and, if it has a mass above the $2 m_t$ threshold, also to top quarks. The relevant fermionic widths are given by
\begin{align}
\label{eq:BRZtt}
\Gamma(Z' \to t\bar t) &\simeq \frac{N_C }{24\pi} \tilde g^2 q'^2 (s_L^4+s_R^4) m_{Z'} , \\
\Gamma(Z' \to \mu^+\mu^-) &\simeq  \frac{ 1 }{12\pi} \tilde g^2 (q'^2_{\mu,V}+q'^2_{\mu,A}) m_{Z'}\,.
\end{align}
neglecting the $m_t^2/m_{Z'}^2$ suppressed terms. Similar expressions apply to potential $Z'\to \nu\bar\nu$ and/or $Z'\to \tau^+\tau^-$ decays, with obvious replacements in the notation. 
For $Z'$ that predominantly couples to one left-handed lepton flavor (two lepton flavors with the same strength) one has ${\rm Br}(Z'\to \ell \bar\ell) \approx 0.5 (0.25)$ for each charged lepton\,.

In Fig. \ref{fig:param_space} we show the constraint from the recent ATLAS high-mass dilepton resonance search~\cite{ATLAS-CONF-2017-027} in the $\tilde g q'$ and $m_Z'$ plane. We fix the heavy $T$ top partner mass to be $m_T=5 m_{Z'}$ with $s_R=0.4$ and taking $q'_{\mu,V}=-q'_{\mu, A}=q'/3$.  The $s_L\sim s_R v/m_T$ is small enough so that electroweak precision tests are not constraining in the shown parameter space. 
For the above parameter choice the branching ratios to $t\bar t$ and $\mu^+\mu^-$ are similar. 
Following ATLAS analysis we use a $40\%$ acceptance for the dominant $Z' j$ production channel and show the bounds derived for $\Gamma_{Z'}/m_{Z'}=0.08$ adjusting for the fact that ATLAS assumes equal decay probabilities for $Z'\to \mu^+\mu^-$ and $Z'\to e^+e^-$. The regions that are excluded by the dilepton resonance search \cite{ATLAS-CONF-2017-027}, for ${\rm Br}(Z'\to \mu\mu)=0.25, 0.50, 1$, are shown in orange.
The $1\sigma$ region preferred by the $b\to s \ell^+\ell^-$ transitions is shown in blue. We see that existing dimuon searches are already covering interesting parameter space. Still, it would be important to gain another order of magnitude in sensitivity of the experimental searches as the precise value of ${\rm Br}(Z'\to \mu\mu)$ is model dependent.  In the most interesting $Z'$ mass range, $m_{Z'}\gtrsim 300$ GeV, the tree level $pp\to t\bar t Z'$ cross section is larger than the loop induced $pp\to Z' j$ process. Thus, searches for di-muon resonances in association with $t\bar t$ can provide an important additional handle on this model.  

An important probe of  $Z'$ coupling to left-handed muons is the neutrino trident production~\cite{Altmannshofer:2014pba}.
The resulting upper bound on $\tilde{g}q^\prime_\mu$ is given by the dashed purple line in Fig.~\ref{fig:param_space}. This is much more constraining than the bounds from LFU violation in leptonic $Z$ couplings, induced at one loop because  the $Z'$ couples to muons but not the electrons~\cite{Altmannshofer:2016brv} (see also~\cite{Efrati:2015eaa}). Finally, since the heavy quark, $T$, or vectorlike leptons, $L$, are charged under both $Z'$ and hypercharge, one expects kinetic mixing between the $Z'$ and the SM $B$ gauge field at the one-loop level,  $\epsilon\sim 10^{-3}$. This is below present bounds in our preferred range of $Z'$ masses; {see e.g.}~\cite{Curtin:2014cca}. 

{\bf Beyond the minimal model.}
The above minimal model can be extended in several ways. For $b\to s\ell^+\ell^-$ decays the only essential ingredient is that the $Z'$ couples to top quarks and to muons. It is very easy to deviate from this minimal assignment, and also couple $Z'$  to $\tau$ leptons without significantly changing the phenomenology. The main effect is on $Z'$ searches since in that case the branching ratio for $Z'\to \mu^+\mu^-$ is reduced, making the searches less sensitive, while on the other hand opening a new search channel of $Z'\to \tau^+\tau^-$. 

The simplest model can also be viewed as a simplified model for strongly interacting NP.  In this case the $Z'$ is the lightest resonance in the strongly interacting sector, while the  $\Phi$ field can be thought of as a condensate of the strong dynamics that breaks dynamically the hidden $U(1)'$ corresponding to the $Z'$ vector. The couplings of $Z'$ to tops and muons then depend on the compositeness fractions of these two fermions.  
It is then also natural for the $Z'$ couplings to the lighter quarks to be suppressed, since these are presumably less composite, while one would expect the couplings of $Z'$ to tau leptons and possibly $b$-quarks to be  enhanced. In this case the $Z'\to \mu^+\mu^-$ branching ratio can be significantly smaller than in the minimal renormalizable model we considered above, while searches for resonances in the ditau channel can become more sensitive (see e.g.~\cite{Faroughy:2016osc}). 

{\bf Conclusions.}
In conclusion, we 
introduced a $Z'$ model, whose defining feature is that the $Z'$ couples to the up-sector, and that can explain the $b\to s \mu^+\mu^-$ anomaly. The $V-A$ structure of the quark current in the $b\to s$ transition is a clear prediction of such models. The $b\to s \mu^+\mu^-$ decay is due to a $Z'$ coupling to muons and top quarks, where the flavor changing transition is predominantly due to a top-$W$ penguin loop. The flavor structure is of the minimal flavor violating type, naturally leading to $b\to s \mu^+\mu^-$ decays as the most important precision flavor observables. The $Z'$ is  expected to be light, $m_{Z'}\lesssim$ TeV, and can be as light as a few 100\,GeV. It can be searched for in dimuon and ditop channels, either in inclusive searches or in a production in association with $Z$, or with a $t\bar t$ pair. 

~~

\begin{acknowledgments}
{The authors would like to thank Jorge Martin Camalich for initial involvement with the work, as well as Wolfgang Altmannshofer, Yosef Nir and Jesse Thaler for useful discussions and comments on the manuscript. 
We thank Paddy Fox, Ian Low and Yue Zhang for pointing out a mistake in Eq.~\eqref{eq:BRZtt} and the corresponding effects in the $Z^\prime$ production processes at the LHC.
We would also like to thank the organizers of the workshop ``Portoro\v z 2017: New physics at the junction of flavor and collider phenomenology'' for the stimulating environment facilitating the completion of this work.  
J.F.K. acknowledges the financial support from the Slovenian Research Agency (research core funding No. P1-0035 and J1-8137). 
The work of Y.S.  is supported by the U.S. Department of Energy (DOE) under Grants Nos. DE-SC-00012567 and No. DE-SC-00015476. 
J.Z. is supported in part by the U.S. National Science Foundation under CAREER Grant PHY-1151392 and by the DOE grant DE-SC0011784.  \\}
\end{acknowledgments}

\bibliography{cksz_ref}

\end{document}